\documentclass[twocolumn,prl,A4paper,ColorBG]{revtex4}

\usepackage[final]{graphics}
\usepackage{amssymb}
\usepackage{amsmath}
\usepackage{latexsym}
\usepackage{pstricks,pst-node,pst-text,pst-3d}
\usepackage{pslatex}
\begin{document}


\title{Manning condensation in two dimensions}
\author{Yoram Burak}
\email{yorambu@kitp.ucsb.edu}
\affiliation{Kavli Institute for Theoretical Physics, UCSB,
Santa Barbara, California 93106, USA}
\author{Henri Orland}
\affiliation{Service de Physique Th\'{e}orique, CE-Saclay,
91191 Gif sur Yvette, France}
\date{May, 2005}

\begin{abstract}
We consider a macroion confined to a cylindrical cell and
neutralized by oppositely charged counterions. 
Exact results are obtained for the two-dimensional version
of this problem, in which ion-ion and ion-macroion interactions
are logarithmic. In particular, the threshold for counterion 
condensation is found to be the same as predicted by 
mean-field
theory. With further increase of the macroion charge, 
a series of single-ion condensation transitions takes place.
Our analytical results are expected to be exact in the
vicinity of these transitions
and are in very good agreement
with recent Monte-Carlo
simulation data.
\\
\end{abstract}

\maketitle

Properties of charged polymers in solution are intimately 
related to the distribution of small ions around them.
A key theoretical model for studying this 
distribution
is that of an infinite charged cylinder, 
immersed in a solution containing counterions, and confined
to a cylindrical cell of finite size. 
When
the cell size increases to infinity
only some of the counterions 
remain bound at a finite distance from the cylinder. The 
remaining ions escape to infinity, leaving behind a distribution
of ions that compensates only part of the cylinder's charge.
Furthermore,
below a critical linear charge density (or, equivalently,
above a critical temperature), all the
counterions escape to infinity:
The ion density at any finite distance from the 
charged cylinder is zero.

The existence of a critical temperature, above which all ions
decondense
is predicted within mean-field (MF) theory 
\cite{FuossKatchalsky,LeBretZimm}.
We consider here the equivalent two-dimensional
(2d) problem where ion-ion interactions, as well as
ion-macroion interactions, are logarithmic. 
Some thermodynamic properties can 
be evaluated exactly in this case, without resorting
to the MF approximation. In particular, we find that
the decondensation temperature is the same as that 
predicted by MF
theory -- suggesting that a similar conclusion might hold in
the 3d case, where 
ion correlation effects are expected to be weaker than in 2d.
The same conclusion was pointed out very recently
by MC simulations in 2d and in 3d \cite{NajiNetz05}, 
in which no deviation from the MF 
decondensation temperature was found numerically.

We begin by briefly discussing the MF theory for
a charged cylinder of radius $a$ in 3d, 
confined in a cylindrical
cell of radius $R$. The
MF electrostatic potential depends only
on the radial coordinate $r$ and obeys the Poisson-Boltzmann
equation \cite{AndelmanReview}
\begin{equation}
-\frac{1}{4\pi} {\bf \nabla}^2\phi = \tilde{\lambda}\theta(\tilde{r})
{\rm e}^{-\phi} - \frac{\xi}{2\pi}\delta(\tilde{r}-1) 
\label{eq:PBeq}
\end{equation}
in which ${\bf r}$, the spatial coordinate, 
was rescaled by the cylinder radius:
${\bf \tilde{r}} = {\bf r}/a$, 
and $\phi$ is the reduced electrostatic potential,
in units of the thermal energy $k T$. 
We assume that counterions carry a positive charge
$e$ and the cylinder is negatively charged, 
with a linear charge density $-e \rho$. 
This charge density enters 
Eq.~(\ref{eq:PBeq}) via $\xi = l_B \rho$, the so-called
Manning parameter \cite{Manning1}, 
and $l_B = e^2/k T$ is the Bjerrum length.
The step function $\theta(\tilde{r})$ is equal to 
unity for $1 < \tilde{r} < R/a$ and to zero elsewhere, and
the boundary condition, $\phi'(\tilde{r} = R/a) = 0$, enforces
charge neutrality. Finally,
$\tilde{\lambda}$ is a rescaled fugacity, which does not have any
physical consequence since changing its value merely shifts the 
MF solution $\phi$ by
a constant. The only dimensionless parameters in the problem are thus
$\xi$ and $R/a$. 

By defining $u = {\rm log}(\tilde{r}) = {\rm log}(r/a)$ and 
$\varphi = \phi - 2u$, Eq.~(\ref{eq:PBeq}) becomes
\begin{equation}
-\frac{1}{4\pi}\frac{{\rm d}^2 \varphi}{{\rm d}u^2} = 
\tilde{\lambda}{\rm e}^{-\varphi}
\label{eq:MFu}
\end{equation}
for $0 \leq u \leq L$ with boundary conditions
\begin{equation}
\left.\frac{{\rm d}\varphi}{{\rm d}u}\right|_{u = 0} = 2(\xi-1)
\ , \ 
\left.\frac{{\rm d}\varphi}{{\rm d}u}\right|_{u = L} = -2 \ ,
\label{eq:MFub}
\end{equation}
where $L = {\rm log}(R/a)$. 
Equations (\ref{eq:MFu})--(\ref{eq:MFub}) 
can be interpreted as describing an ionic solution
confined between two parallel planar 
surfaces -- one at $u = 0$, another at $u = L$, having
surface charges
\begin{equation}
\left.\sigma\right|_{u = 0} = -\frac{1}{2\pi}(\xi-1) \ \ \ ; \ \ \ 
\left.\sigma\right|_{u = L} = -\frac{1}{2\pi}
\label{eq:MFcharges}
\end{equation}
(using units such that $l_B = 1$.) In this equivalent planar problem,
the surface at $u = L$ is negatively charged and thus
always attracts the positively charged counterions. On the other
hand, the surface at $u = 0$ may be positively or negatively
charged, depending on $\xi$: For $\xi < 1$ ions are repelled
from the positively charged surface, and escape to infinity
as $L \rightarrow \infty$; For $\xi > 1$
a finite fraction of the ions remain bound, so as to neutralize the
negatively charged surface at $u = 0$. 

The mapping from cylindrical geometry to a planar one
provides an instructive way to understand the behavior of the
MF solution \cite{FuossKatchalsky}, but is valid only on the MF level.
On the other hand, in the 2d case
we show that a similar transformation is exact,
on the Hamiltonian level.

We begin with the Hamiltonian
$
{\cal H}_n = 2 q q' \sum_{i=1}^{n} {\rm log}(r_i/a) - q'^2 \sum_{i \neq j}{\rm log}
|{\bf r}_i-{\bf r}_j|,
$
which describes $n$ point-like 
ions of charge q' interacting with a central disc of charge q
and radius $a$ in 2d.
By analogy with the 3d model, we assume that ions 
are confined to the 
radial coordinates $a \leq r \leq R$. Charge neutrality  
requires $q/q' =  n$ so that, in contrast to the 3d case, 
the number of ions
is finite. The partition function
is given by $Z_n = (1/n!) \prod_i \int_a^R {\rm d}r_i r_i 
\int_0^{2 \pi}{\rm d}\theta_i \ {\rm exp}(-\beta H)$ 
where $\beta = 1/kT$. We define 
a Manning parameter $\xi = \beta q q'$ so that
\begin{equation}
\beta {\cal H}_n = 2\xi \sum_{i = 1}^{n}{\rm log}(r_i/a) - \frac{\xi}{n} \sum_{i \neq j}
{\rm log}|{\bf r}_i-{\bf r}_j|,
\label{eq:Hn}
\end{equation}
where charge-neutrality is assumed. For convenience, in the following we set $\beta = 1$.

In the grand-canonical ensemble, the partition function can be transformed into a 
field-theory form (as outlined in Ref.~\cite{NetzOrland00})
\begin{eqnarray}
& & Z_{\Lambda} = \sum_{n = 0}^{\infty}\frac{\lambda^n}{n!}{\rm exp}(- {\cal H}_n)  
\propto \int{\cal D}\varphi\,{\rm exp}\left\{-\frac{1}{ q'^2}\times \right. \\
& & \left. \int{\rm d}{\bf \tilde{r}}
\left[\frac{1}{8\pi}({\bf \nabla}\varphi)^2 - i\varphi \frac{\xi}{2\pi}
\delta(\tilde{r}-1) - \tilde{\lambda} \theta(\tilde{r}){\rm exp}(-i\varphi)\right]\right\}.
\nonumber
\end{eqnarray}
As $q'^2 \rightarrow 0$ the prefactor inside the exponential tends
to infinity. Hence, MF theory [Eq.~(\ref{eq:PBeq})] becomes exact, for
any fixed value of $\xi$, in the
thermodynamic limit $n \rightarrow \infty$.
In the following,
we analyze the canonical partition function $Z_n$ for finite $n$,
characterized by the two parameters $\xi$ and $n$ (or, alternatively, $q$ and
$q'$).
To proceed, we note that
\begin{equation}
Z_n = \frac{\zeta}{n!}\int{\rm d}u_i\int{\rm d}\theta_i\,
{\rm exp}(-\tilde{\cal H}),
\label{eq:part}
\end{equation}
where $u_i = {\rm log}(r_i/a)$,
$\zeta = {\rm exp}\left[(n-1)\xi{\rm log}a\right]$,
\begin{equation}
\tilde{H} = \left(\xi -2 +\xi/n\right)\sum_i u_i
- \frac{ q'^2}{2}\sum_{i \neq j} 
v(u_i-u_j,\theta_i-\theta_j),
\label{eq:H1d}
\end{equation}
and
\begin{equation}
v(u,\theta) = -{\rm log}\left[ 2{\rm cosh}u-2{\rm cos}\theta\right].
\label{eq:potential}
\end{equation}
The potential $v$ is linear for $|u| \gg 1$
being then equal, approximately, to $-|u|$. We note that,
since $0 \leq \theta < 2\pi$ is a compact coordinate,
the correction to this linear potential is
short-ranged. 

A charge
$q$, evenly smeared over the $\theta$ interval, 
exerts an exactly linear potential:
\begin{equation}
\frac{-q}{2\pi}\int_0^{2\pi}{\rm d}\theta {\rm log}
\left[2{\rm cosh}u-2{\rm cos}\theta\right] = -q|u| .
\end{equation}
It is thus convenient to interpret the linear term in 
(\ref{eq:H1d}) as
coming from an interaction of the ions with a smeared charge $q_0$ at
$u = 0$ and a smeared charge $q_1$ at $u = L$, which
requires $- q'(q_0-q_1) = \xi-2+\xi/n$.
Adding the same constant to $q_0$ and $q_1$ does not influence 
the force exerted on the ions, and we are free to chose this additive
constant such that the system is overall charge-neutral,
in the following sense: $nq' = -(q_0+q_1)$. With this requirement
there is a unique choice of $q_0$ and $q_1$:
\begin{equation}
q'q_0 = -\xi+1-\frac{\xi}{2n} \ \ ;\ \  
q'q_1 = -1 + \frac{\xi}{2n} \ ,
\end{equation}
which bears some resemblance to Eq.~(\ref{eq:MFcharges}).

So far,
we made an exact transformation of the 
problem from cylindrical coordinates into a problem defined
on a strip: 
the coordinate $u$ goes from 0 to $L$,
and the coordinate $\theta$ is periodic (see Fig.~1). On the ($u$,$\theta$)
strip, ions
interact with each other through a potential of the form $-q'^2|u_1-u_2|$, 
augmented by a short-range contribution. 
They also interact with 
two smeared charges, $q_0$ at $u = 0$
and $q_1$ at $u = L$, and the system is overall charge-neutral.
Note that any critical property
of the system should be exactly captured by the long-range 
linear potential term.

\begin{figure}
\scalebox{0.35}{\includegraphics{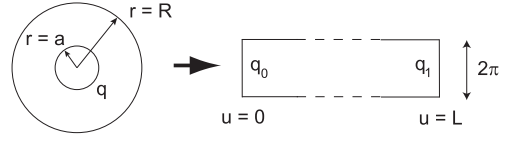}}
\caption{The transformation from cylindrical
geometry to a problem defined on the ($u$,$\theta$) strip
(schematic representation).
}
\end{figure}

Since we are interested in the behavior when $L \rightarrow \infty$,
we next introduce an approximation, treating
the 2d strip as a one-dimensional (1d) domain, with a purely linear
ion-ion interaction.
This can be thought of as the result of 
coarse-graining on a scale of order $2\pi$.
Scaling analysis of the partition function shows that in the 1d model,
the value of some observables is the same, when $L \rightarrow 
\infty$,
as in the 2d problem -- for example, the number of ions
between $u = 0$ and $u = \alpha L$, for any $0 \le \alpha \le 1$. 
Therefore we expect the number of bound ions, evaluated
in the 1d approximation, to be the same as in the 2d problem
\footnote{The ion-ion potential, 
Eq.~(\ref{eq:potential}), is 
divergent when $u_1 = u_2$ and $\theta_1 = \theta_2$,
but this does not lead to a divergence in the partition function
since the corresponding Boltzmann weight vanishes}.

In the 1d model, the partition function is 
$Z_{1d} = (1/n!)\prod_{i = 1}^n\int_0^L{\rm d}x_i\,
{\rm exp}(-H_{1d})$ where
\begin{equation}
H_{1d} = \frac{1}{2}\int_0^L{\rm d}x\,q(x)\psi(x) - \frac{1}{4}
\int_0^L{\rm d}x\left(\frac{{\rm d}\psi}{{\rm d}x}\right)^2,
\end{equation}
$q(x)$ is the one-dimensional 
charge density, including the boundary charges at $0$ and $L$,
and ${\rm d}^2\psi/{\rm d}x^2 = -2q(x)$. Charge
neutrality ensures that ${\rm d}\psi/{\rm d}x = 0$ outside
the interval [0,L]. 
To evaluate $Z_{1d}$, the $n$
particles can be ordered according to their position 
(canceling the $1/n!$ in $Z_{1d}$). The derivative 
${\rm d}\psi/{\rm d}x$ is then
equal to $-2q_0$ between 0 and $x_1$ and decreases in a stepwise
fashion by $2q'$ at each ion position $x_i$,
%
%
so that
\begin{eqnarray}
\label{eq:Z1dintegral}
Z_{1d} & = & \int_0^L{\rm d}x_1\int_{x_1}^L{\rm d}x_2
\cdots\int_{x_{n-1}}^L{\rm d}x_n \times \\
& & {\rm exp}\left[-\alpha_0 x_1 -\alpha_1(x_2-x_1) \cdots
- \alpha_n(L-x_n)\right] ,\nonumber
\end{eqnarray}
where $\alpha_i = (q_0 + i q')^2$.
Note that this expression could have also been obtained directly by
writing the partition function (\ref{eq:part}),
with a linear electrostatic potential  $v(x,\theta) \approx -|x|$
obtained from
(\ref{eq:potential})
in the limit $L \to \infty$.

It follows from Eq.~(\ref{eq:Z1dintegral}) that 
$Z_{1d} = f_0 \circ f_1\circ \cdots \circ f_n(L)$
is the convolution of $f_0,\cdots,f_n$, evaluated at $u = L$, where
$f_i(u) = {\rm exp}(-\alpha_i u)$ are defined for
$u \ge 0$. The Laplace transform of $Z_{1d}(L)$ is
thus
\begin{equation}
{\cal Z}_{1d}(s) = \prod_{k = 0}^n\frac{1}{\alpha_k+s},
\end{equation}
so that, performing the inverse Laplace transform,
\begin{equation}
Z_{1d} = \sum_{k = 0}^{n} c_k {\rm exp}(-\alpha_k L) \ \ ; \ \ 
c_k = \prod_{j \neq k}\frac{1}{\alpha_j-\alpha_k}.
\label{eq:Z1d}
\end{equation}
Note that neither $\alpha_k$ or $c_k$ depend on $L$.

In the limit $L\rightarrow \infty$,
$Z_{1d}$ is dominated by the term $k = k^*$ having the smallest $\alpha_{k}$.
When the Manning parameter $\xi = 0$,
this dominating term is $k^* = 0$; with increase of $\xi$,
$k^*$ increases in a stepwise fashion, changing
by unity at $n$ threshold values (where
$\alpha_{k} = \alpha_{k-1}$), 
\begin{equation}
\xi_k = \frac{n}{n+1-k}.
\label{eq:xik}
\end{equation}
Each one of these discontinuities in $k^*$
corresponds to a thermodynamic transition. In the following, we
analyze the
behavior of several quantities at these transitions. More details
will be presented in a separate publication.

To evaluate the contact density $n(0)$, it is sufficient to consider
the distribution function of $x_1$ (an ion at $x = 0$ is
necessarily the closest to the origin)
\begin{equation}
n_1(x_1) = (\alpha_0-\alpha_{k^*}){\rm e}^{-(\alpha_0-\alpha_{k^*})x_1}.
\label{eq:x1dist}
\end{equation}
We thus find that $n(0) = \alpha_0-\alpha_{k^*}$ is
equal to
\begin{equation}
n(0) = k^*\left[-2+\left(2-\frac{k^*-1}{n}\right)\xi\right],
\label{eq:nzero}
\end{equation}
where
\begin{equation}
k^* = \left\{
\begin{array}{cll}
0 & , & \xi < 1 \\
\left\lfloor 1+n\left(1-\xi^{-1}\right) \right\rfloor
&,& \xi \ge 1
\end{array}
\right. .
\label{eq:kstar}
\end{equation}
Below the first threshold at $\xi_1 = 1$, $n(0)$
vanishes, whereas above this threshold it is finite.
Therefore the threshold for ion condensation is 
the same as predicted by MF theory. Note that
the contact density is continuous at $\xi = \xi_1$. 
This is true also at each one of the other transitions $\xi_k$.
However the derivative of $n(0)$ with respect to $\xi$ is
discontinuous.

In the original, cylindrical problem, our result for $n(0)$
translates into an ion concentration
$\rho(a) = 1/(2\pi a^2) n(0)$. In Fig. 2 we compare 
this result with $a^2 \rho(a)$, as obtained
from MC simulation of the full 2d problem.
Although we used the approximate 1d model, 
the agreement between the analytical prediction and 
simulation is very good. 
In the limit $n\rightarrow\infty$, 
the contact density
approaches the MF theory prediction,
$\rho(a) \rightarrow (\xi-1)^2/(2\pi a^2 \xi)$.

\begin{figure}
\scalebox{0.30}{\includegraphics{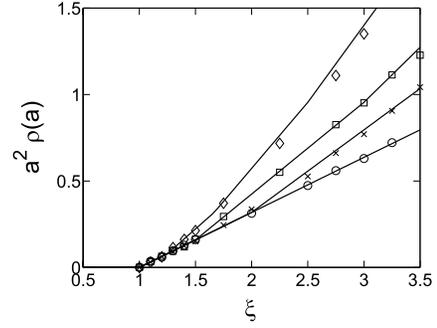}}
\caption{The contact density, $a^2 \rho(a)$, as obtained from 
Eq.~(\ref{eq:nzero}) (lines), compared with MC simulation
results \cite{NajiNetzPrivate} 
($L = 300$): $n = 1$ (circles), 2 (crosses), 3 (squares),
and 5 (diamonds). When $\xi < 1$ the contact density vanishes.}
\end{figure}

An exact sum rule, similar to the contact theorem for the
planar electric double layer \cite{Contact}, 
relates the contact density
in the 2d strip to the number of bound ions:
$n(0) = q_0^2-(q_0+k^*q')^2$ [in agreement with
Eq.~(\ref{eq:nzero})].
This relation is obtained by comparing the pressure across
the plane $u = 0$ to the pressure acting across a 
plane $u = u_0$, 
where $u_0$ is chosen to be far away from both $u = 0$ and
$u = L$.
Since the sum rule is exact in both the 2d problem
and the 1d approximation, equality in the number of bound
ions implies that $\rho(a)$ in the cylindrical problem, 
as calculated from Eq.~(\ref{eq:nzero}), is exact.

To evaluate the density at $u > 0$, the distribution of all
ions $x_2, x_3, \ldots$ must be evaluated. 
We find that $k^*$ [Eq.~(\ref{eq:kstar})] 
is equal to the number of bound ions, and that for
the $m$-th bound ion, the Laplace transform of the
distribution function is
\begin{equation}
{\cal N}_m(s) = \prod_{j = 1}^{m}\frac{\gamma_j}{s+\gamma_j} ,
\end{equation}
where
\begin{equation}
\gamma_m = (k^*-m+1)\left[-2 + \frac{2(n+1)-m-k^*}{n}\xi\right].
\end{equation}
%
The Laplace transform of the total particle density is thus
\begin{equation}
{\cal N}(s) = \frac{\gamma_1}{s+\gamma_1}\left[
1 + \frac{\gamma_2}{s+\gamma_2}\left[1+\ldots
\left[1 + \frac{\gamma_{k^*}}{s+\gamma_{k^*}}\right]
\ldots\right]\right].
\end{equation}
This result provides a particularly simple expression for all
moments of the single-ion distribution in the cylindrical 
coordinates, because the ($-k$)-th moment,
\begin{eqnarray}
\label{eq:moments}
\langle r^{-k}\rangle 
& = &
\frac{2\pi}{n}\int_a^{\infty}r{\rm d}r\, r^{-k}\rho(r) \\
& = & 
\frac{a^{-k}}{n}\int_0^{\infty}{\rm d}u\,{\rm exp}(-k u) n(u) 
 = \frac{a^{-k}}{n}{\cal N}(k) ,
\nonumber
\end{eqnarray}
where $\rho(r) = n(u)/(2\pi r^2)$ is the ion density in the
cylindrical coordinates. Figure 3 shows
$(a/\xi) \langle 1/r\rangle$, obtained from
Eq.~(\ref{eq:moments}) (solid lines). Rescaling with $\mu = a/\xi$ is 
used to facilitate 
comparison with MC simulation results from Ref.~\cite{NajiNetz05}
(symbols). 
The agreement 
is good but not perfect -- deviations reflect the effect
of using the 1d model as an approximation to the $2d$ model on the
strip. As should be expected, agreement is perfect in the MF 
limit ($n \rightarrow \infty$) as well as in the opposite limit, $n = 1$.

\begin{figure}
\scalebox{0.35}{\includegraphics{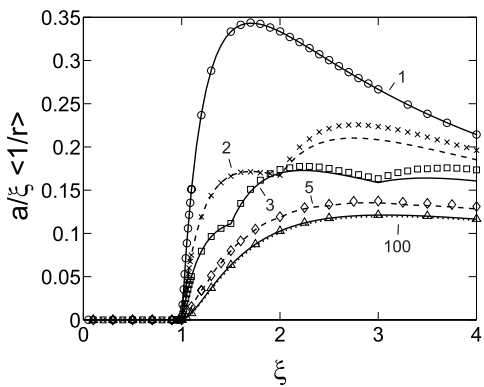}}
\caption{$(a/\xi)\langle r^{-1}\rangle$ as calculated from Eq.~(\ref{eq:moments}) 
for $n = 1$,
2,3,5, and 100 (alternating solid and dashed lines). Symbols 
show MC simulation results \cite{NajiNetz05}
for the same quantity (L = 300). 
The dotted line shows the prediction of MF theory.
}
\end{figure}

%
\begin{figure}[t]
\scalebox{0.35}{\includegraphics{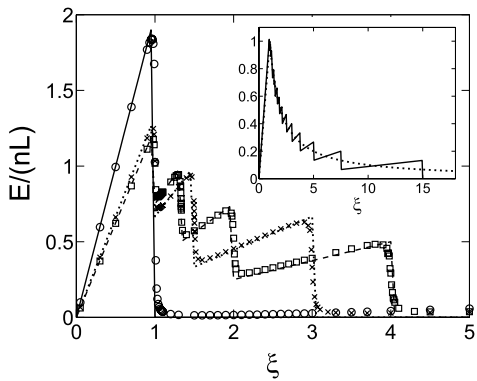}}
\caption {Leading (large $L$) term in $E/(nL)$
[Eq.~(\ref{eq:Energy}), lines], 
compared with MC simulation data from Ref.~\cite{NajiNetz05}
($L = 300$, symbols):
solid line and circles, $n = 1$; dotted line and crosses, 
$n = 2$; dashed line and
squares, $n = 4$. The inset shows $E/(nL)$ for $n = 15$
(solid line), together 
with the MF prediction $E/(nL) = \xi^{-1}$ for $\xi > 1$
(dotted line).
}
\end{figure}

We next evaluate the electrostatic energy.
To leading order in $L$, the free energy in the 1d model is 
$-{\rm log}Z_{1d} = \alpha_{k^*}L$, which corresponds
to a free energy
$(\alpha_k^*-q_1^2)L$ in the original cylindrical problem. 
The mean electrostatic
energy $E$ is found by taking a derivative
$\xi ({\rm d}/{\rm d} \xi)$, which 
is equivalent to $\beta ({\rm d}/{\rm d}\beta)$
by virtue of Eq.~(\ref{eq:Hn}), and yields
\begin{equation}
E = \frac{\xi}{n}(n-k^*)(n-k^*+1)L.
\label{eq:Energy}
\end{equation}
One may expect to find $E = q'^2 (n-k^*)^2 L$, the 
electrostatic energy of a cylindrical capacitor having 
charges $\pm q' (n-k^*)$ on its inner and outer surfaces.
Equation (\ref{eq:Energy}) is similar to this expression, but
there is a correction (second parentheses,
third term), whose contribution goes to zero only in 
the limit of large $n$. We expect Eq.~(\ref{eq:Energy}) to
be the exact leading term in the electrostatic energy
for $L\rightarrow\infty$. A comparison with MC simulation
data (L = 300) \cite{NajiNetz05} is shown in Fig.~4.

%
%
Finally, close to each one of the transition points $\xi = \xi_k$,
$\alpha_k$ approaches $\alpha_{k-1}$ and, as seen from
Eq.~(\ref{eq:Z1d}), both
$c_k$ and $c_{k-1}$ diverge. Concentrating only on their
divergent contribution to the free energy,
we find that
\begin{equation}
E \simeq E_0 + \frac{\xi}{|\xi-\xi_k|}
\end{equation}
on both sides of the transition, where $E_0$ is the 
leading term in L [Eq.~(\ref{eq:Energy})]. 
The leading divergence in 
the heat capacity $\partial E/\partial T$ 
follows as $\xi_k^2/(\xi-\xi_k)^2$. Scaling arguments,
previously presented in Ref.~\cite{NajiNetz05}, are thus
in agreement with the analytical result.

In summary, the counterion condensation problem in 2d
is treated here analytically, taking ion-ion correlations into
account. A series of single-ion
condensation transitions is found with increasing $\xi$, 
in agreement with recent MC 
simulations \cite{NajiNetz05}, the first of these occurring at the MF theory
transition, $\xi = 1$. A possible experimental realization of this
problem may be obtained with parallel, rod-like polyelectrolytes.
Being an analogue of the 3d problem with lower dimensionality,
the 2d model suggests that the Manning transition temperature 
in 3d is exact
even in the presence of ion-ion correlations.

We acknowledge discussions with A. Naji and R.~R Netz,
and thank them for sharing with us their simulation data
prior to publication.
This research was supported in part by the National Science
Foundation under Grant No. PHY99-07949.


\end{document}